\documentclass[conference]{IEEEtran}

\usepackage{cite}
\usepackage{amsmath,amssymb,amsfonts}
 
\usepackage{graphicx}
\usepackage{textcomp}
\usepackage{xcolor}
\usepackage{epsfig}
\usepackage{float}
\usepackage{bm}
\usepackage{graphicx}
\usepackage{amsfonts,amssymb}
\usepackage{booktabs}
\usepackage{array}
\usepackage{tabularx}
\usepackage{multirow}
\usepackage{subfigure}
\usepackage{color}
\usepackage{epstopdf}
\usepackage{makecell}
\usepackage{url}
\usepackage{gensymb}
\usepackage{amsmath}
\usepackage{xcolor}
\usepackage[numbers,sort&compress]{natbib}
\usepackage{algorithmicx}
\usepackage{algorithm}

\usepackage{algpseudocode}
 
\def\BibTeX{{\rm B\kern-.05em{\sc i\kern-.025em b}\kern-.08em
		T\kern-.1667em\lower.7ex\hbox{E}\kern-.125emX}}
\begin{document} 
	
	\title{
		Revisiting Pre-analysis Information Based Rate Control in x265
		 \vspace{-.5em}}

	%\author{\IEEEauthorblockN{Hewei Liu}}
 
	%\and
	%\IEEEauthorblockN{Shuyuan Zhu}
	%\IEEEauthorblockA{\textit{University of Electronic Science and Technology of China}\\
	%Chengdu, China}
	%\and
	%\IEEEauthorblockN{Ruiqin Xiong}
	%\IEEEauthorblockA{\textit{Peking University}\\
	%Beijing, China}
	%\and
	%\IEEEauthorblockN{Bing Zeng}
	%\IEEEauthorblockA{\textit{University of Electronic Science and Technology of China}\\
	%Chengdu, China}
	%}
	%\author{Hewei Liu$^{\ast}$,  \\
		
	%	{\begin{minipage}{\linewidth}\begin{center}
	%				\begin{tabular}{ccc}
	%					$^{\ast}$University of Electronic Science and Technology of China, Chengdu, China\\
	%					%$^{\dag}$Peking University, Beijing, China\\
	%				\end{tabular}
	%	\end{center}\end{minipage}}
	%	\vspace{-1em}}
	
	%\name{Hewei Liu$^{\ast}$, Shuyuan Zhu$^{\ast}$, Ruiqin Xiong$^{\dagger}$, and Bing Zeng$^{\ast}$}
	%\address{$^{\ast}$University of Electronic Science and Technology of China, Chengdu, China\\
	%$^{\dagger}$ Peking University, Beijing, China}
	
	\maketitle
	
	\begin{abstract}
		Due to the excellent compression and high real-time performance, x265 is widely used in practical applications. Combined with CU-tree based pre-analysis, x265 rate control can obtain high rate-distortion (R-D) performance. However, the pre-analysis information is not fully utilized, and the accuracy of rate control is not satisfactory in x265 because of an empirical linear model. In this paper, we propose an improved cost-guided rate control scheme for x265. Firstly, the pre-analysis information is further used to refine the bit allocation. Secondly, CU-tree is combined with the $\lambda$-domain model for more accurate rate control and higher R-D performance.  Experimental results show that compared with the original x265, our method can achieve 10.3\% BD-rate gain with only 0.22\% bitrate error.
	\end{abstract}
	
	\begin{IEEEkeywords}
		Rate control, pre-analysis, x265.
	\end{IEEEkeywords}
	
	\vspace{-0.5em}
	\section{Introduction}
 	Rate control plays an important part in practical video coding systems, e.g. in real-time communication applications. The rate control aims to minimize the distortion under the rate constraint.
 	
 	Traditional rate control algorithms build on the $Q$-domain \cite{qmodel}, which enables rate control by determining quantization parameter (QP) and has been integrated in the AVC reference software. Later, Li et al. \cite{lambdamodel} proposed a $\lambda$-domain method to improve the accuracy of rate control, which is adopted in the HEVC reference software. In \cite{lsx}, a recursive Taylor expansion (RTE) method is proposed to further improve the rate control accuracy. Nonetheless, the above algorithms hardly consider temporal dependence, which undoubtedly helps to improve the performance. \cite{motiones1} and \cite{motiones2} establish a temporal propagation chain by simulating the motion estimation in the pre-analysis process. However, an obvious increase of complexity is inevitable owing to motion estimation based on the original resolution.
 	
 	As far as practical encoders are concerned, pre-analysis based on downsampled pictures is usually preferred for a limited complexity increase. A macroblock-tree (MB-tree) scheme~\cite{mbtree} using downsampled pictures is adopted in the x264 encoder~\cite{x264}. By means of pre-analysis, propagation cost and intra cost are obtained and QP offset is generated in MB-level. MB-tree was inherited by the HEVC encoder x265~\cite{x265} dubbed as CU-tree and x265 inherited the excellent performance of MB-tree as well. Due to its high performance, x265 has been widely used in practical applications such as FFmpeg~\cite{ffmpeg}, VLC media player~\cite{VLC}, and HandBrake~\cite{hand}.
 	
 	Although CU-tree achieves great BD-rate gain, the pre-analysis information can be further exploited. In~\cite{lzy}, Liu et al. proposed to improve the frame-level based QP by pre-analysis information. However, it may still be vulnerable to bitrate inaccuracy due to the adopted traditional linear model. In this paper, we further exploit pre-analysis information to refine the bit allocation and meanwhile combine CU-tree with $\lambda$-domain by conditionally increasing the frame-level QP. Experimental results show that the proposed method greatly improves x265 rate control accuracy and R-D performance.  

 	The rest of this  paper is organized as follows. Section \ref{Sec.bg} describes the background of CU-tree and introduces the theory of the $\lambda$-domain model. Section \ref{Sec.ours} presents the proposed strategy of conditional QP increase and the bit allocation strategy. Experimental results are presented in section \ref{Sec.exp}. Finally, conclusions are drawn in Section  \ref{Sec.con}.
	\section{Background} \label{Sec.bg}

	\subsection{The CU-Tree Scheme In x265}\label{Sec.cutree}
	CU-tree calculates propagation cost and intra cost for each CU and generates a QP offset to imply different block significance. CU-tree is implemented with 8$\times$8 CU as a basic unit and source pictures are downsampled to accelerate this process. For each 8$\times$8 CU, it operates as follows:
 	
 	1$).$ Estimate sum of absolute Hadamard transform  difference (SATD) cost of intra mode (denoted as $C_{intra}$) and inter mode (denoted as $C_{inter}$). It should be noted that $C_{inter}$ is forced to be less than or equal to $C_{intra}$. $C_{propagate}$ is used to estimate all of the information contributing to future pictures, and it is set to zero for the first running frame because no information has been collected.
 	
 	2$).$ $F_{p}$, which represents the fraction of information from a CU to be propagated to its reference frames, is calculated as
 	\begin{eqnarray} 
 	 F_{p} = \frac{C_{intra} - C_{inter}}{C_{intra}}.
 	\end{eqnarray} 
 	
 	3$).$ The total amount of information that relies on current CU is $(C_{intra}+C_{propagate})$, which indicates all the future dependencies in the lookhead range. The information $AC_{propagate}$ propagated to the CU's reference block is estimated as
 	\begin{eqnarray} 
 	\begin{array}{l}
 	AC_{propagate} = (C_{intra}+ C_{propagate})\cdot F_{p}.
 	\end{array}
 	\end{eqnarray} 
 	Note that reference blocks with size 8$\times$8 may overlap with multiple 8$\times$8 CUs. Therefore $AC_{propagate}$ 
 	is distributed to each 8$\times$8 CU by overlap area $area_{overlap}$. The revised $C_{propagate}^{'}$ of 8$\times$8 CU is accumulated by
  	\begin{eqnarray}\label{Eq.pcost}
 	 \begin{array}{l}
 	 C_{propagate}^{'} = C_{propagate}+ \frac{area_{overlap}}{64}\cdot AC_{propagate}.
 	 \end{array}
 	 \end{eqnarray} 
 		
 	4$).$ Finally, $\Delta QP_{CU}$ can be obtained for each CU 
 	\begin{eqnarray}\label{Eq.cqp}
 	\Delta QP_{CU}=-s\cdot {log_2} \left( \frac{C_{intra}+C_{propagate}^{'}}{C_{intra}}
 	\right)
 	\end{eqnarray} 
 	where $s$ is a constant parameter. Actually, CU-tree is consistent with one's intuition, because if one CU contributes more for future blocks, it will have a larger propagate cost, and finally larger QP offset will bring higher reconstruction quality to this CU.

 	\subsection{$\lambda$-Domain Model}\label{Sec.lambda}
 	The cost function of the  $\lambda$-domain model is formulated by
 
 	\begin{eqnarray} \label{Eq.cost}
	\mathop {min}  J = D+\lambda R
 	\end{eqnarray}
	where $R$ and $D$ are the bitrate and distortion respectively for a basic unit (such as GOP, frame, coding tree unit, and so on). By setting the derivative of (\ref{Eq.cost}) with respect to $R$ to zero, we can have
	\begin{eqnarray}\label{Eq.nccost}
   \lambda  =  - \frac{{\partial D}}{{\partial R}}.
	\end{eqnarray} 
	Combining the hyperbolic model $D = C{R^{ - K}}$, it can be further obtained
\begin{eqnarray}\label{Eq.lambda}
	\lambda  = CK{R^{ - K - 1}} \triangleq  \alpha {R^\beta }
\end{eqnarray} 
where $\alpha$ and $\beta$ are both constant parameters. Once the target bit of a basic unit is allocated,  $\lambda$ can be calculated by (\ref{Eq.lambda}). QP is calculated by the QP-$\lambda$ model
\begin{eqnarray}\label{Eq.ql}
QP  = 4.2005\cdot ln{\lambda}+13.7122.
\end{eqnarray} 

After encoding a basic unit, the actual bit allocation (denoted by $R_{a}$)  and the actual $\lambda$ (denoted by $\lambda_{a}$) can be derived. Correspondingly $\alpha$ and $\beta$ in (\ref{Eq.lambda}) will be updated by
\begin{eqnarray}\label{Eq.update}
\begin{array}{l}
\mathop \alpha ' = \alpha  + \delta_\alpha  \cdot (ln\lambda_{a}-ln(\alpha R_a^\beta))\cdot \alpha \\
\beta ' = \beta  + \delta_\beta  \cdot (ln\lambda_{a}-ln(\alpha R_a^\beta))\cdot lnR_a  
\end{array}
\end{eqnarray} 
where $\alpha '$ and $\beta '$ represent updated $\alpha$ and $\beta$. $\delta_\alpha$ and $\delta_\beta$ denote the learning rate, which reflects the update speed.

\section{The Proposed Rate Control Scheme}\label{Sec.ours}
In this section, we propose a cost-guided rate control scheme. The proposed scheme is described on the basis of the typical hierarchical structure of four layers, e.g. as shown in Fig.\ref{gop}, where I frame belongs to layer 0, P frame belongs to layer 1, and B frame can belong to layer 2 or layer 3. It should be noted that B frame of layer 3 is a non-reference frame. Due to the different layer characteristics of various frame types, the scheme is proposed according to the frame type.
\subsection{Scheme For I Frame}\label{Sec.oursi}
I frame is treated as a special GOP with only one frame. To adapt to CU-tree combined with the $\lambda$-domain model, a conditional QP increase strategy is adopted.

Firstly, the  SATD of the original frame is calculated to obtain frame-level $QP_{Ibase}$~\cite{gt}. Then, CU-tree will decrease QP based on $QP_{Ibase}$. In order to improve the rate control accuracy and R-D performance, we utilize the average $\Delta QP_{CU}$ of all 8$\times$8 CUs to increase the QP conditionally. Specifically, when the $j$th frame in the sequence is I frame, its final QP is
\begin{eqnarray} \label{Eq.qpi}
QP_I^j=
\left\{
\begin{array}{rcl}
\begin{aligned}
&QP_{Ibase}, if \, epp_j < T\\
& QP_{Ibase} + avg(abs(\Delta QP_{CU})), otherwise\\
\end{aligned}
\end{array} \right.
\end{eqnarray} 
where $T$ is an empirical threshold, and $epp_j$ can reflect the difference of frames in the lookahead range. $epp_j$ is defined as
\begin{eqnarray} \label{Eq.mae}
epp_j \triangleq \frac{1}{{n - 1}}\sum\limits_{i = j}^{j+n - 2} {\frac{{||{f_i} - {f_{i + 1}}|{|_1}}}{{W_d \cdot H_d}}} 
\end{eqnarray} 
where $n$ denotes the lookahead range, $f_i$ represents the $i$th downsampled frame, $W_d$ and $H_d$ are respectively the width and height of downsampled frames, and $|| \cdot ||_1$ is the L1 norm.
\subsection{Scheme For P/B Frame}\label{Sec.oursbp}

The loss function in a GOP is expressed as
\begin{eqnarray}\label{Eq.flcost}
\begin{array}{l}
\mathop {min}\limits_{{R_1},{R_2},...,{R_M}} D = \sum\limits_{i = 1}^M {{D_i}\,\ s.t} \sum\limits_{i = 1}^M {{R_i} \le \,{R_{gop}}} 
\end{array}
\end{eqnarray}
where $R_i$ and $D_i$ are respectively the bitrate and distortion for the $i$th frame, $M$ represents the total number of frames in a GOP, and $R_{gop}$ denotes the target bits in a GOP. Converting (\ref{Eq.flcost}) to an unconstrained problem, we can have
\begin{eqnarray}\label{Eq.ucflcost}
\begin{array}{l}
\mathop {min}\limits_{{R_1},{R_2},...,{R_M}} \,\,J = \sum\limits_{i = 1}^M {({D_i} + \lambda_g {R_i})}
\end{array}
\end{eqnarray}
where $\lambda_g$ is the global Lagrange multiplier in a GOP. When encoding the $j$th frame, (\ref{Eq.ucflcost}) can be solved by setting its derivative with respect to $R_j$ to zero 
\begin{eqnarray}\label{Eq.dezero}
\begin{array}{l}
\frac{{\partial J}}{{\partial {R_j}}} = \frac{{\partial \sum\nolimits_{i = 1}^M {{D_i}} }}{{\partial {R_j}}} + \lambda_g  = 0.
\end{array}
\end{eqnarray}
Substituting  (\ref{Eq.nccost}) into  (\ref{Eq.dezero}), we can further have 
\begin{eqnarray}\label{Eq.de1}
\begin{array}{l}
\begin{aligned}	 
\frac{{\partial \sum\nolimits_{i = 1}^M {{D_i}} }}{{\partial {R_j}}} \cdot \frac{{\partial {R_j}}}{{\partial {D_j}}} &=  - \lambda_g  \cdot \frac{{\partial {R_j}}}{{\partial {D_j}}}\\
1 + \frac{{\partial \sum\nolimits_{i = j + 1}^M {{D_i}} }}{{\partial {D_j}}} &=  - \lambda_g  \cdot \frac{{\partial {R_j}}}{{\partial {D_j}}}
\end{aligned}	 
\end{array}
\end{eqnarray}
let $\frac{1}{w_{layer}}$ denote $\frac{{\partial \sum\nolimits_{i = j + 1}^M {{D_i}} }}{{\partial {D_j}}} + 1$. (\ref{Eq.de1}) can be rewritten as
\begin{eqnarray}\label{Eq.de33}
\begin{array}{l}
\begin{aligned}
{\lambda _j} &= -\frac{{\partial D_j}}{{\partial {R_j}}}={w_{layer}}\lambda_g
\end{aligned}	 
\end{array}
\end{eqnarray}
from which it can be seen that the Lagrange multiplier of each frame is a scaled global Lagrange multiplier. Besides, (\ref{Eq.de33}) indicates if one frame is important (i.e., greatly influences the quality of future frames), the small $w_{layer}$ is desired to allocate more bits and obtain higher R-D performance for this frame. 

Considering the same $w_{layer}$ is shared by frames belonging to the same layer and frames of layer 3 are non-reference frames, without loss of generality, $w_3$ is set to 1, $w_2$ for layer 2 is set as 1$/$1.2599~\cite{zm}, and $w_1$ for layer 1 is set to $1/(1+k)$
\begin{eqnarray}\label{Eq.w}
\begin{array}{l}
{w_{layer}} = \left\{ {\begin{array}{*{20}{c}}
	{1,}&{layer = 3}\\
	{1/1.2599,}&{layer = 2}\\
	{1/(1 + k),}&{layer = 1}
	\end{array}}. \right.
\end{array}
\end{eqnarray}
In (\ref{Eq.w}), $k$ is defined as
\begin{eqnarray}\label{Eq.k}
\begin{array}{l}
		%R_{cost} =\frac{C_{propagate}}{C_{intra}}\\\\
	k = \frac{avg(C_{propagate}^{'})}{avg(C_{intra})} \cdot \frac{{QP_{last}^2}}{c} \cdot ln(QP_{last}) 
\end{array}
\end{eqnarray}
where $avg(C_{propagate}^{'})$ and $avg(C_{intra})$ are the average values of $C_{propagate}^{'}$ and $C_{intra}$ in the frame respectively. $QP_{last}$ denotes the QP of the colocated frame in the last GOP, and $c$ is a constant parameter.
Combining (\ref{Eq.de33}), (\ref{Eq.w}) and (\ref{Eq.lambda}), it can be obtained
\begin{eqnarray} \label{Eq.bi}
\begin{array}{l}
\begin{aligned}
 \sum\limits_{i = j}^M {{R_i}} =\sum\limits_{i = j}^M {{{\left( {\frac{{{w_{layer}}\lambda_g }}{{{\alpha _i}}}} \right)}^{\frac{1}{{{\beta _i}}}}}}  &= {R_{gopleft}}
 
\end{aligned}
\end{array}
\end{eqnarray}
%.where $j$ represents the current frame is the $j$th frame (encoding order) in the GOP
where the bitrate is summed from the $j$th frame, $M$ is the total number of frames in the GOP, and $R_{gopleft}$ represents the left bits in the GOP. In (\ref{Eq.bi}), only $\lambda_g$ is undetermined and it is solved by the bisection method in this paper. Then $QP_{PBbase}$ of the $j$th frame can be calculated according to (\ref{Eq.ql}). Considering the effects of CU-tree, the final QP of the $j$th frame, which is P or B frame, is
\begin{eqnarray} \label{Eq.qpbp}
\begin{small}
QP_{PB}^j=
\left\{
\begin{array}{rcl}
\begin{aligned}
&QP_{PBbase}, if \, epp_j < T\\
& QP_{PBbase} + avg(abs(\Delta QP_{CU})), otherwise.
\end{aligned}
\end{array} \right.
\end{small}
\end{eqnarray} 
Note that if the $j$th frame is a non-reference frame, $QP_{PB}^j$ is equal to $QP_{PBbase}$. 

After encoding the $j$th frame, update $R_{gopleft}$ in a GOP, and update $\alpha$, $\beta$ by (\ref{Eq.update}). Then set $i=j+1$ to calculate  the Lagrange multiplier for the next frame.
\begin{figure} [t]
	\centering
	\centerline{\epsfig{figure=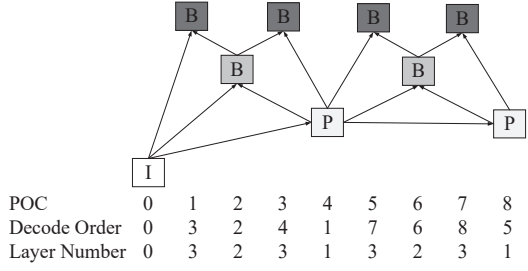,height=3.5cm,width=7cm}}
	\caption{The 4-layer hierarchical structure.}\label{gop}
\end{figure}

For easy understanding, the proposed rate control algorithm is summarized in Algorithm 1.

\begin{algorithm}[t]
	\caption{The proposed rate control algorithm}%标题
	\label{alg1}%标签
	\begin{algorithmic}[1]
		\renewcommand{\algorithmicrequire}{ \textbf{Input:}}  
		  {\Require  
		Total number of frames $f$, GOP size $M$ and bitrate left in a GOP $R_{gopleft}$.}
	\renewcommand{\algorithmicrequire}{ \textbf{Output:}}  
	{\Require  
		The frame-level QP of the frame.}
		\State Initialization: $i=0$.
		\While{$i$\textless  $f$}  
		\If{the $i$th frame is I frame} 
		\State{Calculate $avg(abs(\Delta QP_{CU}))$ of the frame.}
		\State Calculate $epp_i$ of the frame by (\ref{Eq.mae}).
		\State Use the SATD to calculate the $QP_{Ibase}$ \cite{gt}.  
		\State Calculate $QP_{I}^i$ by (\ref{Eq.qpi}) and encode the frame.
		\State After encoding the frame, update $\alpha$, $\beta$ by (\ref{Eq.update}).
		\State $i = i + 1$.
		\Else    
		\For{$j = i$ : $max(i+M-1, f-1)$}
		\If{the $j$th frame is I frame} 
		\State break 
		\EndIf
		\State{Calculate $avg(abs(\Delta QP_{CU}))$ of the frame.}
		\State Calculate $epp_j$ of the frame by (\ref{Eq.mae}).
		\If{$j\textless M$}
		\State set $w$ value according to \cite{lambdamodel}.
		\Else
		\State Set $w$ value by (\ref{Eq.w}).
		\EndIf
		\State Substitute $w$ into (\ref{Eq.bi}) to obtain $\lambda_g$.
		\State Substitute $\lambda_g$ into (\ref{Eq.de33}) to obtain $\lambda_j$.
		\State Substitute $\lambda_j$ to (\ref{Eq.ql}) to calculate $QP_{PBbase}$.
		\State Calculate $QP_{PB}^j$ by (\ref{Eq.qpbp}) and encode the frame.  
		\State After encoding the frame, update the $R_{gopleft}$.
		\State After encoding the frame, update $\alpha$, $\beta$ by (\ref{Eq.update}).
		
		\EndFor
		\State $i = j$.
		\EndIf
		\EndWhile
	\end{algorithmic}
\end{algorithm}
 
	\section{Experimental Results}\label{Sec.exp}
 	We implement the proposed method in x265 version 3.2.1 with a typical hierarchical GOP structure as shown in Fig.\ref{gop}. Sequences recommended by JCT-VC~\cite{jctvc} are tested. Firstly, we encode each sequence with four fixed QPs (QP = 22, 27, 32, and 37). Then, the actual bitrate under each QP is set as the target bitrate for rate control. In the experiment, $c$ in (\ref{Eq.k}) is set to 4791.5, and $T$ in (\ref{Eq.qpi}) is set as 2.5. We do experiment based on 1-pass average bitrate mode and fast presetting in x265. Our rate control method is tested under the single thread (no wpp, no pmode, no pme, no pools, and frame threads is set to 1), and so are the anchors. In x265, YUV-PSNR which is the weighted sum of Y-PSNR, U-PSNR, and V-PSNR (6:1:1) is used as the quality metric. To ensure optimal BD-rate performance in the sense of YUV-PSNR, both our method and anchors close the aq-mode, psy-rd and psy-rdoq.
 	\begin{table}[t]
 		%\vspace{-1.5em}
 		%\renewcommand\arraystretch{1.2}
 		\centering
 		%\fontsize{8}{10}\selectfont
 		\caption{Comparison of BD-rate relative to x265}\label{res}
 		\vspace{-.5em}{
 			\begin{tabular} {c c  c c c}
 				\Xhline{1.2pt}
 				\hline\hline
 				\multicolumn{2}{c}{\multirow{2}{*}{}}&\multicolumn{3}{c }{BD-Rate(\%)} \\
 				%\cline{3-6}
 				\multicolumn{2}{c}{}&Ours &Ours-wo/c &Li \\
 				\hline
			 &BasketballDrill 		&-4.36 	&-1.85 		&-1.64 \\
			&BasketballPass 		&-2.06  &-0.19		&0.87 \\
			&BlowingBubbles 		&-4.08 	&-1.19		&1.24 			\\
			&BQMall 				&-2.66 	&1.35		&2.57 			\\
			&BQSquare 				&-2.74 	&-3.75		&-2.43 			\\
			&BQTerrace 			&-3.11  &-1.39		&0.39		\\
			&Cactus 				&-7.28 	&-4.14		&-3.88			 \\
			&FourPeople 			&-22.70 &-22.85		&-21.06			 \\
			&Johnny 				&-25.36 &-25.36		&-24.24	 			\\
			&KristenAndSara 		&-23.61 &-23.61 	&-21.18 		 \\
			&PartyScene 			&-2.43  &0.31		&3.30			 \\
			&Vidyo1 				&-15.65 &-15.65 	&-14.60 		 \\
			&Vidyo3 				&-14.71 &-16.73 	&-14.76 		 \\
			&Vidyo4 				&-13.48 &-14.32 	&-13.39 		 \\
 				&\textbf{Average}  &\textbf{-10.30} &-9.24 &-7.77   \\
 				\hline\hline
 				\Xhline{1.2pt}
 		\end{tabular}}
 		\label{tab:Margin_settings}
 		\vspace{-1em}
 	\end{table}

  	\begin{table}[t]
 	%\vspace{-1.5em}
 	%\renewcommand\arraystretch{1.2}
 	\centering
 	%\fontsize{8}{10}\selectfont
 	\caption{Comparison of bitrate error}\label{res2}
 	\vspace{-.5em}{
 		\begin{tabular} {c c  c c c c c}
 			\Xhline{1.2pt}
 			\hline\hline
 			\multicolumn{2}{c}{\multirow{2}{*}{}}& \multicolumn{4}{c}{Bitrate Error(\%)}\\
 			%\cline{3-6}
 			\multicolumn{2}{c }{} &Ours &Ours-wo/c &Li &x265 \\
 			\hline
 		&BasketballDrill 			  	&0.002 &0.002	&0.007	&0.187 \\
 		&BasketballPass 				&0.008 &0.006	&0.011	&0.743\\
 		&BlowingBubbles 				&0.047 &0.477 &0.480	&0.428\\
 		&BQMall 						&0.015	&0.245&0.233	&0.143\\
 		&BQSquare 						&0.007 &0.788 &0.839	&0.770\\
 		&BQTerrace 						&0.035	&0.470&0.489	&0.439\\
 		&Cactus 						&0.007 &0.011	&0.016	&0.109 \\
 		&FourPeople 					&0.566 &0.594	&0.587	&0.994 \\
 		&Johnny 				 		&0.786 &0.786	&0.782	&1.359	\\
 		&KristenAndSara 				&0.527 &0.527	&0.404	&0.368 \\
 		&PartyScene 					&0.007 &0.120	&0.091	&0.883 \\
 		&Vidyo1 						&0.670 &0.670	&0.679	&1.003 \\
 		&Vidyo3 						&0.091 &0.711	&0.647	&0.540 \\
 		&Vidyo4 						&0.309 &0.488	&0.418	&0.261 \\
 			&\textbf{Average}  &\textbf{0.220} &0.421 &0.406 &0.588  \\
 			\hline\hline
 			\Xhline{1.2pt}
 	\end{tabular}}
 	\label{tab:Margin_settings}
 	\vspace{-1em}
 \end{table}

 Besides the BD-rate, bitrate error (BE) is defined to evaluate the rate control accuracy
 \begin{eqnarray}  
 \begin{array}{l}
 BE = {\frac{{|R_t - R_a|}}{{R_t}}} \times 1000 $\textperthousand $
 \end{array}
 \end{eqnarray}
  where $R_t$ is the target bitrate of the sequence and $R_a$ is the actual bitrate of the sequence. 
  
  The comparisons of BD-rate and bitrate error are shown in Table \ref{res} and Table \ref{res2} respectively, where Li denotes the proposed method in \cite{ll}, Ours represents the complete proposed method, and Ours-wo/c is the proposed method without conditional QP increase strategy. As seen from the tables, the proposed strategy of conditional QP increase and the bit allocation strategy are both effective, and the complete proposed method can achieve 10.3\% BD-rate gain with only 0.22\% bitrate error which is superior than the anchors both in the rate control accuracy and the R-D performance.

 \begin{figure} [t]
 	 	\begin{minipage}[a]{0.48\linewidth}
 		\centering
 		\centerline{\epsfig{figure=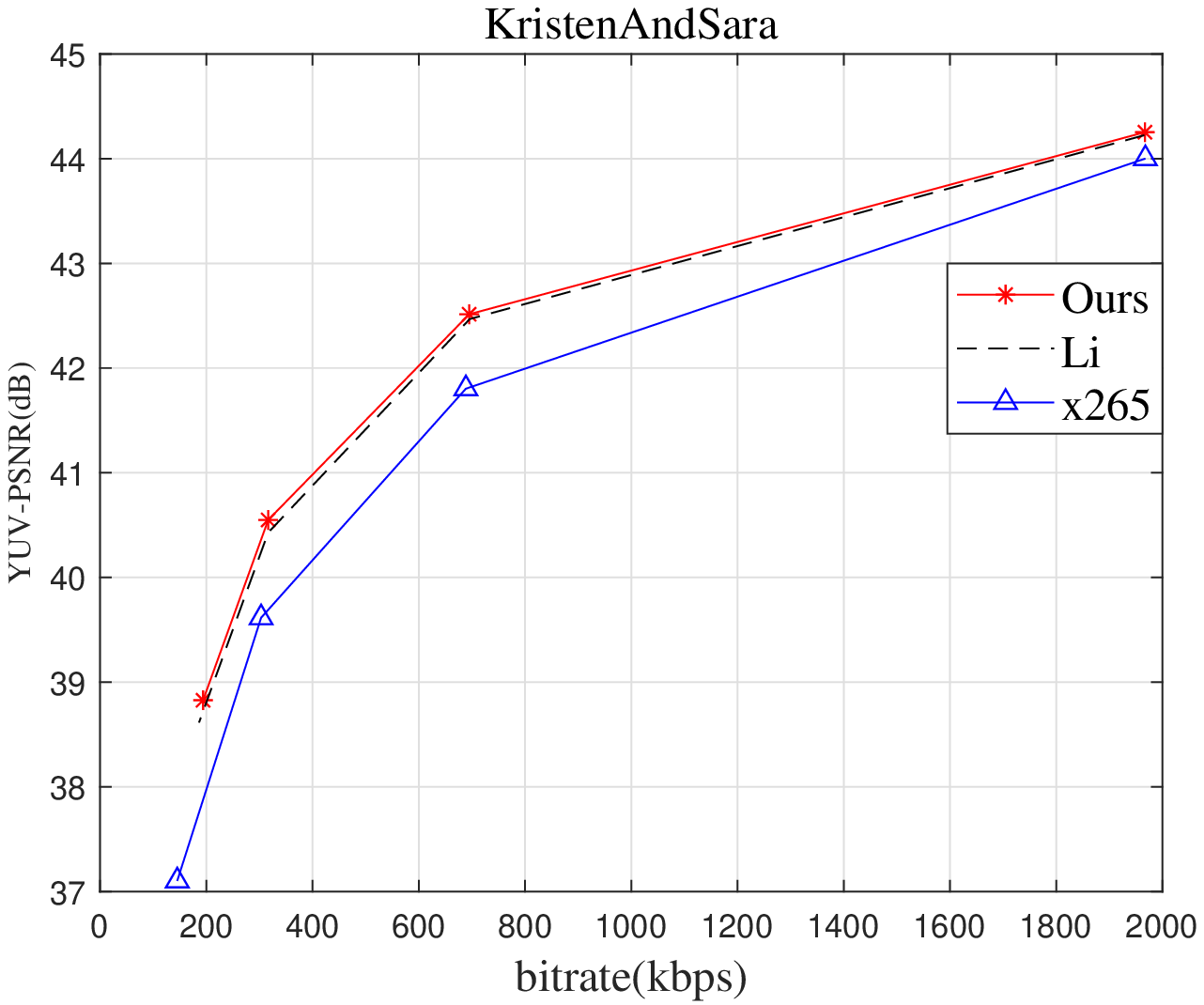,height=3.8cm,width=4.2cm}}
 		\centerline{\small(a)}\medskip
 	\end{minipage}
 	\begin{minipage}[a]{0.5\linewidth}
 		\centering
 		\centerline{\epsfig{figure=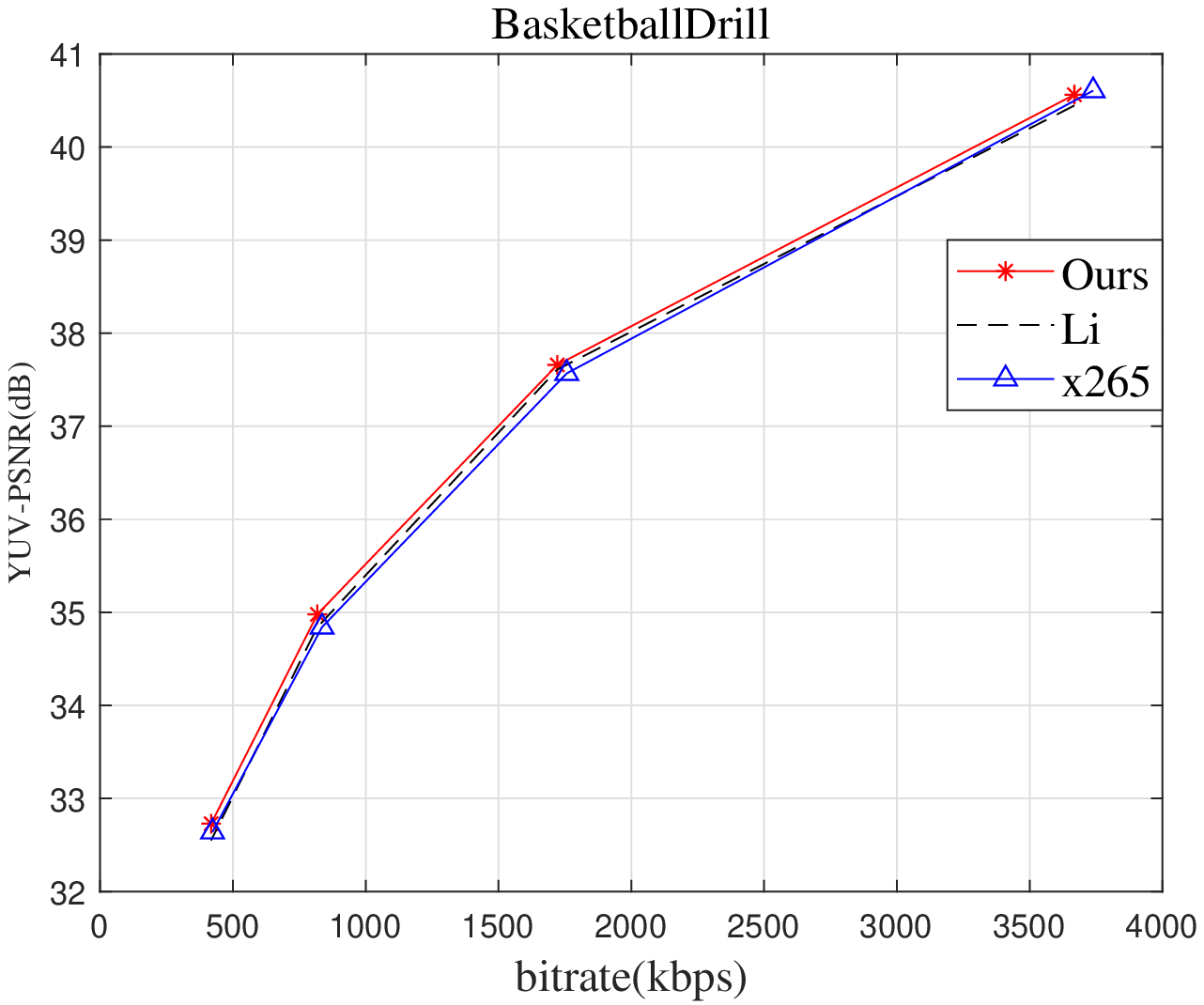,height=3.8cm,width=4.2cm}}
 		\centerline{\small(b)}\medskip
 	\end{minipage}
% 	\begin{minipage}[a]{0.48\linewidth}
% 		\centering
% 		\centerline{\epsfig{figure=Figures/fig4.eps,height=3cm,width=3.8cm}}
% 		\centerline{\small(c)}\medskip
% 	\end{minipage}
% 	\begin{minipage}[a]{0.52\linewidth}
% 		\centering
% 		\centerline{\epsfig{figure=Figures/fig5.eps,height=3cm,width=3.8cm}}
% 		\centerline{\small(d)}\medskip
% 	\end{minipage}
 \small{\caption{R-D curve comparisons.}\label{rd}}
 \end{figure}
 
  In addition, Fig.\ref{rd} shows the R-D curve comparisons of two sequences (KristenAndSara and BasketaballDrill) which also indicate excellent R-D performance of the proposed rate control algorithm.

	\section{Concluding Remarks}\label{Sec.con}
	In this paper, we propose a cost-guided rate control method for x265. Firstly, the bit allocation strategy is refined by further exploiting pre-analysis information; Secondly, the conditional QP increase strategy is applied to further improve the rate control accuracy and R-D performance. Compared with the original x265, our scheme can achieve 10.3\% BD-rate gain with only 0.22\% bitrate error on average.

\end{document}